# Cost Underestimation in Public Works Projects: Error or Lie?

By Bent Flyvbjerg, Mette Skamris Holm, and Søren Buhl



## Abstract


This article presents results from the first statistically significant study of cost escalation in transportation infrastructure projects. Based on a sample of 258 transportation infrastructure projects worth $90 billion (U.S.), it is found with overwhelming statistical significance that the cost estimates used to decide whether important infrastructure should be built are highly and systematically misleading. The result is continuous cost escalation of billions of dollars. The sample used in the study is the largest of its kind, allowing for the first time statistically valid conclusions regarding questions of cost underestimation and escalation for different project types, different geographical regions, and different historical periods. Four kinds of explanation of cost underestimation are examined: technical, economic, psychological, and political. Underestimation cannot be explained by error and is best explained by strategic misrepresentation, i.e., lying. The policy implications are clear: In debates and decision making on whether important transportation infrastructure should be built, those legislators, administrators, investors, media representatives, and members of the public who value honest numbers should not trust the cost estimates and cost-benefit analyses produced by project promoters and their analysts.




**Introduction**

Existing comparative studies of actual and estimated costs in transportation infrastructure development are few. Where such studies exist, they are typically single-case studies or they cover a sample of projects too small to allow systematic, statistical analyses (Bruzelius et al., 1998; Fouracre et al., 1990; Hall, 1980; Nijkamp & Ubbels, 1999; Pickrell, 1990; Skamris & Flyvbjerg, 1997; Szyliowicz & Goetz, 1995; Walmsley & Pickett, 1992). To our knowledge, only one study exists that, with a sample of 66 transportation projects, approaches a large-sample study and takes a first step toward valid statistical analysis (Merewitz, 1973a, 1973b).[1] Despite their many merits in other respects, these studies have not produced statistically valid answers regarding the question of whether one can trust the cost estimates used by decision makers and investors in deciding whether or not to build new transportation infrastructure. Because of the small and uneven samples used in existing studies, different studies even point in opposite directions, and researchers consequently disagree regarding the credibility of cost estimates. Pickrell (1990), for instance, concludes that cost estimates are highly inaccurate, with actual costs being typically much higher than estimated costs, while Nijkamp and Ubbels (1999) claim that cost estimates are rather correct. Below we will see who is right.

The objective of the study reported here is to answer the following questions in a statistically valid manner: How common and how large are differences between actual and estimated costs in transportation infrastructure projects? Are the differences significant? Are they simply random errors? Or is there a statistical pattern to the differences that suggests other explanations? What are the implications for policy and decision making regarding transportation infrastructure development?



**Four Steps To Understanding Deceptive Cost Estimation**

We see four steps in the evolution of a body of scholarly research aimed at understanding practices of cost underestimation and deception in decision making for transportation infrastructure. The first step was taken by Pickrell (1990) and Fouracre, Allport, and Thomson (1990), who provided sound evidence for a small number of urban rail projects that substantial cost underestimation is a problem, and who implied that such underestimation may be caused by deception on the part of project promoters and forecasters. The second step was taken by Wachs (1990), who established--again for a small sample of urban rail projects--that lying, understood as intentional deception, is, in fact, an important cause of cost underestimation. Wachs began the difficult task of charting who does the lying, why it occurs, what the ethical implications are, etc.

The problem with the research in the first two steps is that it is based on too few cases to be statistically significant; the pattern found may be due to random properties of the small samples involved. This problem is solved in the third step, taken with the work reported in this article. Based on a large sample of transportation infrastructure projects, we show that (1) the pattern of cost underestimation uncovered by Pickrell and others is of general import and is statistically significant, and (2) the pattern holds for different project types, different geographical regions, and different historical periods. We also show that the large-sample pattern of cost underestimation uncovered by us lends statistical support to the conclusions about lying and cost underestimation arrived at by Wachs for his small sample.

The fourth and final step in understanding cost underestimation and deception would be to do for a large sample of different transportation infrastructure projects what Wachs did for his small sample of urban rail projects: establish whether systematic deception actually takes place, who does the deception, why it occurs, etc. This may be done by having a large number of forecasters and project promoters, representing a large number of projects, directly



express, in interviews or surveys, their intentions with and reasons for underestimating costs. This is a key topic for further research.

In sum, then, we do not claim with this article to have provided final proof that lying is the main cause of cost underestimation in transportation infrastructure projects. We claim, however, to have taken one significant step in a cumulative research process for testing whether this is the case by establishing the best and largest set of data about cost underestimation in transportation infrastructure planning so far seen, by carrying out the first statistically significant study of the issues involved, and by establishing that our data support and give statistical significance to theses about lying developed in other research for smaller, nonsignificant samples.

As part of further developing our understanding of cost underestimation, it would also be interesting to study the differences between projects that are approved on a competitive basis, by voters at an election, and those that are funded through formula-based allocations. One may speculate that there is an obvious incentive to make a project look better, and hence to underestimate costs, in the campaign leading up to an election. A good single-case study of this is Kain's (1990) article about a rail transit project in Dallas. Votes are cast more often for large rail, bridge, and tunnel projects than for road projects. For example, most U.S. highway funds are distributed to states based on a formula (i.e., there is no competitive process). A state department of transportation (DOT) is likely to have a fixed annual budget for construction. The DOT leadership would presumably want fairly accurate cost estimates before allocating the budget. One may speculate that large cost underestimation is less likely in this situation. There are exceptions to this scenario. Sometimes DOT officials want to persuade state legislators to increase their budget. And states occasionally submit bond issue proposals to voters. In Europe, the situation is similar on important points, although differences also exist. This may explain the result found below, that cost underestimation is substantially lower for roads than for rail, bridges, and tunnels, and that



this is the case both in the U.S. and Europe. Needless to say, more research is necessary to substantiate this observation.

Finally we want to emphasize that although the project sample used in this study is the largest of its kind, it is still too small to allow more than a few subdivisions, if comparative statistical analyses must still be possible. Therefore, in further work on understanding cost underestimation, the sample should be enlarged to better represent different types of projects and different geographical locations. As to project types, data for more private projects would be particularly useful in allowing statistically valid comparisons between public and private sector projects. Such comparisons do not exist today, and nobody knows whether private projects perform better or worse than public ones regarding cost underestimation. The sample should also be enlarged to contain data for more fixed links and rail projects. Such data would allow a better (i.e., a statistically corroborated) comparative understanding of cost underestimation for more specific subtypes of projects like bridges, tunnels, high-speed rail, urban rail, and conventional rail. Such an understanding is non-existent today. As to geography, immediate rewards would be gained from data for projects outside Europe and North America, especially for fixed links and roads. But even for Europe and North America, data on more projects are needed to allow better comparative analysis.

**Measuring Cost Inaccuracy**

The methods used in our study are described in the Appendix. All costs are construction costs. We follow international convention and measure the inaccuracy of cost estimates as so-called "cost escalation" (often also called "cost overrun"; i.e., actual costs minus estimated costs in percent of estimated costs). Actual costs are defined as real, accounted construction costs determined at the time of project completion. Estimated costs are defined as budgeted, or forecasted, construction costs at the time of decision to build. Although the project planning process varies with project type, country, and time, it is typically possible for a



given project to identify a specific point in the process as the time of decision to build.

Usually a cost estimate was available at this point in time for the decision makers. If not, then

the closest available estimate was used, typically a later estimate resulting in a conservative

bias in our measure for inaccuracy (see the Appendix). All costs are calculated in fixed prices

in Euros by using the appropriate historical, sectoral, and geographical indices for discounting

and the appropriate exchange rates for conversion between currencies.

Project promoters and their analysts sometimes object to this way of measuring cost

inaccuracy (Flyvbjerg et al., in press). Various cost estimates are made at different stages of

the process: project planning, decision to build, tendering, contracting, and later

renegotiations. Cost estimates at each successive stage typically progress toward a smaller

number of options, greater detail of designs, greater accuracy of quantities, and better

information about unit price. Thus, cost estimates become more accurate over time, and the

cost estimate at the time of making the decision to build is far from final. It is only to be

expected, therefore, that such an early estimate would be highly inaccurate. And this estimate

would be unfair as the basis for assessing the accuracy of cost forecasting, or so the objection

against using the time-of-decision-to-build estimate goes (Simon, 1991). We defend this

method, however, because when the focus is on decision making, and hence on the accuracy

of the information available to decision makers, then it is *exactly* the cost estimate at the time

of making the decision to build that is of primary interest. Otherwise it would be impossible

to evaluate whether decisions are informed or not. Estimates made after the decision to build

are by definition irrelevant to this decision. Whatever the reasons are for cost increases after

decision makers give the go-ahead to build a project, or however large such increases are,

legislators and citizens--or private investors in the case of privately funded projects--are

entitled to know the uncertainty of budgets. Otherwise transparency and accountability suffer.

We furthermore observe that if the inaccuracy of early cost estimates were simply a matter of

incomplete information and inherent difficulties in predicting a distant future, as project



promoters often say it is, then we would expect inaccuracies to be random or close to random. Inaccuracies, however, have a striking and highly interesting bias, as we will see below.

Another objection to using cost at the time of decision to build as a basis of comparison is that this supposedly would entail the classical error of comparing apples and oranges. Projects change over the planning and implementation process. When, for instance, the physical configuration of the original Los Angeles Blue Line Light Rail project was altered at substantial cost to comprise grade-crossing improvements, upgrading of adjacent streets, better sidewalks, new fences, etc., the project was no longer the same. It was, instead, a new and safer project, and comparing the costs of this project with the costs of the older, less safe one would supposedly entail the apples-and-oranges error. A problem with this argument is that existing research indicates that project promoters routinely ignore, hide, or otherwise leave out important project costs and risks in order to make total costs appear low (Flyvbjerg et al., in press; Wachs, 1989, 1990). For instance, environmental and safety concerns may initially be ignored, even though they will have to be taken into account later in the project cycle if the project lives on, and the project is more likely to live on if environmental and safety concerns are initially ignored. Similarly, ignoring or underplaying geological risk may be helpful in getting projects approved, and no other risk is more likely to boomerang back and haunt projects during construction. "Salami tactics," is the popular name used to describe the practice of introducing project components and risks one slice at a time in order to make costs appear low as long as possible. If such tactics are indeed a main mechanism in cost underestimation, as existing research indicates, then, clearly, comparing actual project costs with estimated costs at the time of decision to build does not entail the error of comparing apples and oranges but is simply a way of tracking how what was said to be a small, inexpensive apple turned out to actually be a big, expensive one.

Finally, we observe that if we were to follow the objections against using the cost estimate at the time of decision to build as the basis of tracking cost escalation, it would be



impossible to make meaningful comparisons of costs because no common standard of comparison would be available. We also observe that this method is the international standard for measuring inaccuracy of cost estimates (Fouracre et al., 1990; Leavitt et al., 1993; National Audit Office & Department of Transport 1992; Nijkamp & Ubbels, 1999; Pickrell, 1990; Walmsley & Pickett, 1992; World Bank, 1994). This standard conveniently allows meaningful and consistent comparisons within individual projects and across projects, project types, and geographical areas. This standard, then, is employed below to measure the inaccuracy of cost estimates in 258 transportation infrastructure projects worth $90 billion (U.S.).

**Inaccuracy of Cost Estimates**

Figure 1 shows a histogram with the distribution of inaccuracies of cost estimates. If errors in estimating costs were small, the histogram would be narrowly concentrated around zero. If errors in overestimating costs were of the same size and frequency as errors in underestimating costs, the histogram would be symmetrically distributed around zero. Neither is the case. We make the following observations regarding the distribution of inaccuracies of construction cost estimates:

- Costs are underestimated in almost 9 out of 10 projects. For a randomly selected project, the likelihood of actual costs being larger than estimated costs is 86%. The likelihood of actual costs being lower than or equal to estimated costs is 14%.

- Actual costs are on average 28% higher than estimated costs (sd=39).

- We reject with overwhelming significance the thesis that the error of overestimating costs is as common as the error of underestimating costs ($p<0.001$; two-sided test, using the binomial distribution). Estimated costs are biased, and the bias is caused by systematic underestimation.



- We reject with overwhelming significance the thesis that the numerical size of the error of underestimating costs is the same as the numerical size of the error of overestimating costs ($p < 0.001$; non-parametric Mann-Whitney test). Costs are not only underestimated much more often than they are overestimated or correct, costs that have been underestimated are also wrong by a substantially larger margin than costs that have been overestimated.

We conclude that the error of underestimating costs is significantly much more common and much larger than the error of overestimating costs. Underestimation of costs at the time of decision to build is the rule rather than the exception for transportation infrastructure projects. Frequent and substantial cost escalation is the result.

[Figure 1 app. here]

**Cost Underestimation by Project Type**

In this section, we test whether different types of projects perform differently with respect to cost underestimation. Figure 2 shows histograms with inaccuracies of cost estimates for each of the following project types: (1) rail (high-speed; urban; and conventional, inter-city rail), (2) fixed link (bridges and tunnels), and (3) road (highways and freeways). Table 1 shows the expected (average) inaccuracy and standard deviation for each type of project.

[Figure 2 & Table 1 about here]

Statistical analyses of the data in Table 1 show both means and standard deviations to be different with a high level of significance. Rail projects incur the highest difference between actual and estimated costs with an average of no less than 44.7%, followed by fixed links averaging 33.8% and roads at 20.4%. An F-test falsifies the null hypothesis at a very high level of statistical significance that type of project has no effect on percentage cost



escalation (p<0.001). Project type matters. The substantial and significant differences between project types indicate that pooling the three types of projects in statistical analyses, as we did above, is strictly not appropriate. Therefore, in the analyses which follow, each type of project will be considered separately.

Based on the available evidence, we conclude that rail promoters appear to be particularly prone to cost underestimation, followed by promoters of fixed links. Promoters of road projects appear to be relatively less inclined to underestimate costs, although actual costs are higher than estimated costs much more often than not for road projects as well.

Further subdivisions of the sample indicate that high-speed rail tops the list of cost underestimation, followed by urban and conventional rail, in that order. Similarly, cost underestimation appears to be larger for tunnels than for bridges. These results suggest that the complexities of technology and geology might have an effect on cost underestimation. These results are not statistically significant, however. Even if the sample is the largest of its kind, it is too small to allow repeated subdivisions and still produce significant results. This problem can only be solved by further data collection from more projects.

We conclude that the question of whether there are significant differences in the practice of cost underestimation among rail, fixed link, and road projects must be answered in the affirmative. The average difference between actual and estimated costs for rail projects is substantially and significantly higher than that for roads, with fixed links in a statistically nonsignificant middle position between rail and road. The average inaccuracy for rail projects is more than twice that for roads, resulting in average cost escalations for rail more than double that for roads. For all three project types, the evidence shows that it is sound advice for policy and decision makers as well as investors, bankers, media, and the public to take any estimate of construction costs with a grain of salt, especially for rail and fixed link projects.


## Cost Underestimation by Geographic Location

In addition to testing whether cost underestimation differs for different kinds of projects, we also tested whether it varies with geographical location among Europe, North America, and "other geographical areas" (a group of 10 developing nations plus Japan). Table 2 shows the difference between actual and estimated costs in these three areas for rail, fixed link, and road projects. There is no indication of statistical interaction between geographical area and type of project. We therefore consider the effects from these variables on cost underestimation separately. For all projects, we find that the difference between geographical areas in terms of underestimation is highly significant ($p<0.001$). Geography matters to cost underestimation.

[Table 2 about here]

If Europe and North America are compared separately, which is compulsory for fixed links and roads because no observations exist for these projects in other geographical areas comparisons can be made by t-tests (as the standard deviations are rather different, the Welch version is used). For fixed link projects, the average difference between actual and estimated costs is 43.4% in Europe versus 25.7% North America, but the difference between the two geographical areas is nonsignificant ($p=0.414$). Given the limited number of observations and the large standard deviations for fixed link projects, we would need to enlarge the sample with more fixed link projects in Europe and North America in order to test whether the differences might be significant for more observations. For rail, the average difference between actual and estimated costs is 34.2% in Europe versus 40.8% in North America. For roads, the similar numbers are 22.4% versus 8.4%. Again, the differences between geographical areas are nonsignificant ($p=0.510$ and $p=0.184$, respectively).

We conclude, accordingly, that the highly significant differences we found above for geographical location come from projects in the "other geographical areas" category. The average difference between actual and estimated costs in this category is a hefty 64.6%.



**Have Forecasters Improved Over Time?**

In the previous two sections, we saw how cost underestimation varies with project type and geography. In this section, we conclude the statistical analyses by studying how underestimation varies over time. We ask and answer the question of whether project promoters and forecasters have become more or less inclined over time to underestimate the costs of transportation infrastructure projects. If underestimation were unintentional and related to lack of experience or faulty methods in estimating and forecasting costs, then, a priori, we would expect underestimation to decrease over time as better methods were developed and more experience gained through the planning and implementation of more infrastructure projects.

Figure 3 shows a plot of the difference between actual and estimated costs against year of decision to build for the 111 projects in the sample for which these data are available. The diagram does not seem to indicate an effect from time on cost underestimation. Statistical analysis corroborate this impression. The null hypothesis that year of decision has no effect on the difference between actual and estimated costs cannot be rejected (p=0.22, F-test). A test using year of completion instead of year of decision (with data for 246 projects) gives a similar result (p=0.28, F-test).

[Figure 3 about here]

We therefore conclude that cost underestimation has not decreased over time. Underestimation today is in the same order of magnitude as it was 10, 30, and 70 years ago. If techniques and skills for estimating and forecasting costs of transportation infrastructure projects have improved over time, this does not show in the data. No learning seems to take place in this important and highly costly sector of public and private decision making. This seems strange and invites speculation that the persistent existence over time, location, and project type of significant and widespread cost underestimation is a sign that an equilibrium



has been reached: Strong incentives and weak disincentives for underestimation may have taught project promoters what there is to learn, namely, that cost underestimation pays off. If this is the case, underestimation must be expected and it must be expected to be intentional. We test such speculation below. Before doing so, we compare cost underestimation in transportation projects with that in other projects.

**Cost Underestimation in Other Infrastructure Projects**

In addition to cost data for transportation infrastructure projects, we have reviewed cost data for several hundred other projects including power plants, dams, water distribution, oil and gas extraction, information technology systems, aerospace systems, and weapons systems (Arditi et al., 1985; Blake et al., 1976; Canaday, 1980; Department of Energy Study Group, 1975; Dlakwa & Culpin, 1990; Fraser, 1990; Hall, 1980; Healey, 1964; Henderson, 1977; Hufschmidt & Gerin, 1970; Merewitz, 1973b; Merrow, 1988; Morris & Hough, 1987; World Bank, 1994, n.d.). The data indicate that other types of projects are at least as, if not more, prone to cost underestimation as are transportation infrastructure projects.

Among the more spectacular examples of cost underestimation are the Sydney Opera House, with actual costs approximately 15 times higher than those projected, and the Concorde supersonic airplane, with cost 12 times higher than predicted (Hall, n.d., p. 3). The data also indicate that cost underestimation for other projects have neither increased nor decreased historically, and that underestimation is common in both first- and third-world countries. When the Suez canal was completed in 1869, actual construction costs were 20 times higher than the earliest estimated costs and three times higher than the cost estimate for the year before construction began. The Panama Canal, which was completed in 1914, had cost escalations in the range of 70 to 200% (Summers, 1967, p. 148).



In sum, the phenomena of cost underestimation and escalation appear to be characteristic not only of transportation projects but of other types of infrastructure projects as well.

**Explanations of Underestimation: Error or Lie?**

Explanations of cost underestimation come in four types: technical, economic, psychological, and political. In this section, we examine which explanations best fit our data.

*Technical Explanations*

Most studies that compare actual and estimated costs of infrastructure projects explain what they call "forecasting errors" in technical terms, i.e., in terms of imperfect techniques, inadequate data, honest mistakes, inherent problems in predicting the future, lack of experience on the part of forecasters, etc. (Ascher, 1978; Flyvbjerg et al., in press; Morris & Hough, 1987; Wachs, 1990). Few would dispute that such factors may be important sources of uncertainty and may result in misleading forecasts. And for small-sample studies, which are typical of this research field, technical explanations have gained credence because samples have been too small to allow tests by statistical methods. However, the data and tests presented above, which come from the first large-sample study in the field, lead us to reject technical explanations of forecasting errors. Such explanations simply do not fit the data.

First, if misleading forecasts were truly caused by technical inadequacies, simple mistakes, and inherent problems with predicting the future, we would expect a less biased distribution of errors in cost estimates around zero. In fact, we have found with overwhelming statistical significance ($p<0.001$) that the distribution of such errors has a nonzero mean. Second, if imperfect techniques, inadequate data, and lack of experience were main explanations of the underestimations, we would expect an improvement in forecasting accuracy over time, since errors and their sources would be recognized and addressed through



the refinement of data collection, forecasting methods, etc. Substantial resources have been spent over several decades on improving data and methods. Still our data show that this has had no effect on the accuracy of forecasts. Technical factors, therefore, do not appear to explain the data. It is not so-called forecasting "errors" or cost "escalation" or their causes that need explaining. It is the fact that in 9 out of 10 cases, costs are underestimated.

We may agree with proponents of technical explanations that it is, for example, impossible to predict for the individual project exactly *which* geological, environmental, or safety problems will appear and make costs soar. But we maintain that it is possible to predict the risk, based on experience from other projects, *that* some such problems will haunt a project and how this will affect costs. We also maintain that such risk can and should be accounted for in forecasts of costs, but typically is not. For technical explanations to be valid, they would have to explain why forecasts are so consistent in ignoring cost risks over time, location, and project type.

### *Economic Explanations*

Economic explanations conceive of cost underestimation in terms of economic rationality. Two types of economic explanation exist, one explains in terms of economic self-interest, the other in terms of the public interest. As regards self-interest, when a project goes forward, it creates work for engineers and construction firms, and many stakeholders make money. If these stakeholders are involved in or indirectly influence the forecasting process, then this may influence outcomes in ways that make it more likely that the project will be built. Having costs underestimated and benefits overestimated would be economically rational for such stakeholders because it would increase the likelihood of revenues and profits. Economic self-interest also exists at the level of cities and states. Here, too, it may explain cost underestimation. Pickrell (1990, 1992) pointed out that transit capital investment projects in the U.S. compete for discretionary grants from a limited federal budget each year. This



creates an incentive for cities to make their projects look better, or else some other city may get the money.

As regards the public interest, project promoters and forecasters may deliberately underestimate costs in order to provide public officials with an incentive to cut costs and thereby to save the public's money. According to this type of explanation, higher cost estimates would be an incentive for wasteful contractors to spend more of the taxpayer's money. Empirical studies have identified promoters and forecasters who say they underestimate costs in this manner and with this purpose, i.e., in order to save public money (Wachs, 1990). The argument has also been adopted by scholars, for instance Merewitz (1973b) who explicitly concludes that "keeping costs low is more important than estimating costs correctly" (p. 280).

Both types of economic explanation account well for the systematic underestimation of costs found in our data. Both depict such underestimation as deliberate, and as economically rational. If we now define a lie in the conventional fashion as making a statement intended to deceive others (Bok, 1979, p. 14; Cliffe et al., 2000, p. 3), we see that deliberate cost underestimation is lying, and we arrive at one of the most basic explanations of lying, and of cost underestimation, that exists: Lying pays off, or at least economic agents believe it does. Moreover, if such lying is done for the public good (e.g., to save taxpayers' money), political theory would classify it in that special category of lying called the "noble lie," the lie motivated by altruism. According to Bok (1979) this is the "most dangerous body of deceit of all" (p. 175).

In the case of cost underestimation in public works projects, proponents of the noble lie overlook an important fact: Their core argument--that tax payers' money is saved by cost underestimation--is seriously flawed. Anyone with even the slightest trust in cost-benefit analysis and welfare economics must reject this argument. Underestimating the costs of a given project leads to a falsely high benefit-cost ratio for that project, which in turn leads to



two problems. First, the project may be started despite the fact that it is not economically

viable. Or, second, it may be started instead of another project that would have yielded higher

returns had the actual costs of both projects been known. Both cases result in the inefficient

use of resources and therefore in waste of taxpayers' money. Thus, for reasons of economic

efficiency alone, the argument that cost underestimation saves money must be rejected;

underestimation is more likely to result in waste of taxpayers' money. But the argument must

also be rejected for ethical and legal reasons. In most democracies, for project promoters and

forecasters to deliberately misinform legislators, administrators, bankers, the public, and the

media would not only be considered unethical but in some instances also illegal, for instance

where civil servants would misinform cabinet members or cabinet members would misinform

the parliament. There is a formal "obligation to truth" built into most democratic constitutions

on this point. This obligation would be violated by deliberate underestimation of costs,

whatever the reasons for underestimation may be. Hence, even though economic explanations

fit the data and help us understand important aspects of cost underestimation, such

explanations cannot be used to justify underestimation.

### Psychological Explanations

Psychological explanations attempt to explain biases in forecasts by a bias in the mental make

up of project promoters and forecasters. Politicians may have a "monument complex,"

engineers like to build things, and local transportation officials sometimes have the mentality

of empire builders. The most common psychological explanation is probably that of

"appraisal optimism." According to this explanation, promoters and forecasters are held to be

overly optimistic about project outcomes in the appraisal phase of projects, that is, when

projects are planned and decided (Fouracre et al., 1990, p. 10; Mackie & Preston, 1998;

Walmsley & Pickett, 1992, p. 11; World Bank, 1994, p. 86). An optimistic cost estimate is

clearly a low one. The existence of appraisal optimism in promoters and forecasters would



consequently result in actual costs being higher than estimated costs. Consequently the existence of appraisal optimism would be able to account, in whole or in part, for the peculiar bias of cost estimates found in our data, where costs are systematically underestimated. Such optimism, and associated cost underestimation, would not be lying, needless to say, because the deception involved is self-deception and therefore not deliberate. Cost underestimation would be error according to this explanation.

There is a problem with psychological explanations, however. Appraisal optimism would be an important and credible explanation of underestimated costs if estimates were produced by inexperienced promoters and forecasters, i.e., persons who were estimating costs for the first or second time and who were thus unknowing about the realities of infrastructure building and were not drawing on the knowledge and skills of more experienced colleagues. Such situations may exist and may explain individual cases of cost underestimation. But given the fact that the human psyche is distinguished by a significant ability to learn from experience, it seems unlikely that promoters and forecasters would continue to make the same mistakes decade after decade instead of learning from their actions. It seems even more unlikely that a whole profession of forecasters and promoters would collectively be subject to such a bias and would not learn over time. Learning would result in the reduction, if not elimination, of appraisal optimism, which would then result in cost estimates becoming more accurate over time. But our data clearly shows that this has not happened.

The profession of forecasters would indeed have to be an optimistic bunch to keep their appraisal optimism up throughout the 70-year period our study covers and not learn that they were deceiving themselves and others by underestimating costs. This would account for the data but is not a credible explanation. As observed elsewhere, the incentive to publish and justify optimistic estimates is very strong, and the penalties for having been overoptimistic are generally insignificant (Davidson & Huot, 1989, p. 137; Flyvbjerg et al., in press). This is a better explanation of the pervasive existence of optimistic estimates than an inherent bias



for optimism in the psyche of promoters and forecasters. And "optimism" calculated on the basis of incentives is not optimism, of course; it is deliberate deception. Therefore, on the basis of our data, we reject appraisal optimism as a primary cause of cost underestimation.

### Political Explanations

Political explanations construe cost underestimation in terms of interests and power (Flyvbjerg, 1998). Surprisingly little work has been done that explains the pattern of misleading forecasts in such terms (Wachs, 1990, p. 145). A key question for political explanations is whether forecasts are intentionally biased to serve the interests of project promoters in getting projects started. This question again raises the difficult issue of lying. Questions of lying are notoriously hard to answer, because in order to establish whether lying has taken place, one must know the intentions of actors. For legal, economic, moral, and other reasons, if promoters and forecasters have intentionally fabricated a deceptive cost estimate for a project to get it started they are unlikely to tell researchers or others that this is the case (Flyvbjerg, 1996; Wachs, 1989).

When Eurotunnel, the private company that owns the Channel tunnel, went public in 1987 in order to raise funds for the project, investors were told that building the tunnel was relatively straightforward. Regarding risks of cost escalation, the prospectus read (*The Economist*, 7 October 1989, 37):

> Whilst the undertaking of a tunneling project of this nature necessarily involves certain construction risks, the techniques to be used are well proven#The Directors, having consulted the Mâitre d'Oeuvre, believe that 10%#would be a reasonable allowance for the possible impact of unforeseen circumstances on construction costs.[2] ("Under water, over budget," 1989, p. 37).



Two hundred banks communicated these figures for cost and risk to investors, including a

large number of small investors. As observed by *The Economist* ("Under water, over budget,"

1989), anyone persuaded in this way to buy shares in Eurotunnel in the belief that the cost

estimate was the mean of possible outcomes was, in effect, deceived. The cost estimate of the

prospectus was a best possible outcome, and the deception consisted in making investors

believe in the highly unlikely assumption--disproved in one major construction project after

another--that everything would go according to plan, with no delays; no changes in safety and

environmental performance specifications; no management problems; no problems with

contractual arrangements, new technologies, or geology; no major conflicts; no political

promises not kept; etc. The assumptions were, in other words, those of an ideal world. The

real risks of cost escalation for the Channel tunnel were many times higher than those

communicated to potential investors, as evidenced by the fact that once built, the real costs of

the project were higher by a factor of two compared with forecasts.

Flyvbjerg, Bruzelius, and Rothengatter (in press) document for a large number of

projects that the Everything-Goes-According-to-Plan type of deception used for the Channel

tunnel is common. Such deception is, in fact, so widespread that in a report on infrastructure

and development, the World Bank (1994, pp. ii, 22) found reason to coin a special term for it:

the "EGAP-principle." Cost estimation following the EGAP-principle simply disregards the

risk of cost escalation resulting from delays, accidents, project changes, etc. This is a major

problem in project development and appraisal, according to the World Bank.

It is one thing, however, to point out that investors, public or private, were deceived

in particular cases. It is quite another to get those involved in the deceptions to talk about this

and to possibly admit that deception was intentional, i.e., that it was lying. We are aware of

only one study that actually succeeded in getting those involved in underestimating costs to

talk about such issues (Wachs, 1986, 1989, 1990). Wachs interviewed public officials,

consultants, and planners who had been involved in transit planning cases in the U.S. He



found that a pattern of highly misleading forecasts of costs and patronage could not be explained by technical issues and were best explained by lying. In case after case, planners, engineers, and economists told Wachs that they had had to "cook" forecasts in order to produce numbers that would satisfy their superiors and get projects started, whether or not the numbers could be justified on technical grounds (Wachs, 1990, p. 144). One typical planner admitted that he had repeatedly adjusted the cost figures for a certain project downward and the patronage figures upward to satisfy a local elected official who wanted to maximize the chances of getting the project in question started. Wachs's work is unusually penetrating for a work on forecasting. But, again, it is small-sample research, and Wachs acknowledges that most of his evidence is circumstantial (Wachs, 1986, p. 28). The evidence does not allow conclusions regarding the project population. Nevertheless, based on the strong pattern of misrepresentation and lying found in his case studies, Wachs goes on to hypothesize that the type of abuse he has uncovered is "nearly universal" and that it takes place not only in transit planning but also in other sectors of the economy where forecasting routinely plays an important role in policy debates (Wachs, 1990, p. 146; 1986, p. 28).

Our data give support to Wachs' claim. The pattern of highly underestimated costs is found not only in the small sample of projects Wachs studied; the pattern is statistically significant and holds for the project population mean (i.e., for the majority of transportation infrastructure projects). However, on one point Wachs (1986) seems to conclude somewhat stronger than is warranted: "[F]orecasted costs always seem to be *lower* than actual costs" (p. 24) he says (emphasis in original). Our data show that although "always" (100%) may cover the small sample of projects Wachs chose to study, when the sample is enlarged by a factor of 20-30 to a more representative one, "only" in 86% of all cases are forecasted costs lower than actual costs. Such trifles--14 percentage points--apart, the pattern identified by Wachs is a general one and his explanation of cost underestimation in terms of lying to get projects started fit our data particularly well. Of the existing explanations of cost development in



transportation infrastructure projects, we therefore opt for political and economic explanations. The use of deception and lying as tactics in power struggles aimed at getting projects started and at making a profit appear to best explain why costs are highly and systematically underestimated in transportation infrastructure projects.

**Summary and Conclusions**

The main findings from the study reported in this article all highly significant, and most likely conservative are as follows:

- In 9 out of 10 transportation infrastructure projects, costs are underestimated.

- For rail projects, actual costs are on average 45% higher than estimated costs (sd=38).

- For fixed link projects (tunnels and bridges), actual costs are on average 34% higher than estimated costs (sd=62).

- For road projects, actual costs are on average 20% higher than estimated costs (sd=30).

- For all project types, actual costs are on average 28% higher than estimated costs (sd=39).

- Cost underestimation exists across 20 nations and 5 continents; it appears to be a global phenomenon.

- Cost underestimation appears to be more pronounced in developing nations than in North America and Europe (data for rail projects only).

- Cost underestimation has not decreased over the past 70 years. No learning that would improve cost estimate accuray seems to take place.

- Cost underestimation cannot be explained by error and seems to be best explained by strategic misrepresentation, i.e., lying.



- Transportation infrastructure projects do not appear to be more prone to cost underestimation than are other types of large projects.

We conclude that the cost estimates used in public debates, media coverage, and decision making for transportation infrastructure development are highly, systematically, and significantly deceptive. So are the cost-benefit analyses into which cost estimates are routinely fed to calculate the viability and ranking of projects. The misrepresentation of costs is likely to lead to the misallocation of scarce resources, which, in turn, will produce losers among those financing and using infrastructure, be they tax payers or private investors.

We emphasize that these conclusions should not be interpreted as an attack on public (vs. private) spending on infrastructure, since the data are insufficient to decide whether private projects perform better or worse than public ones as regards cost underestimation. Nor do the conclusions warrant an attack on spending on transportation vs. spending on other projects, since other projects appear to be as liable to cost underestimation and escalation as are transportation projects. With transportation projects as an in-depth case study, the conclusions simply establish that significant cost underestimation is a widespread practice in project development and implementation, and that this practice forms a substantial barrier to the effective allocation of scarce resources for building important infrastructure.

The key policy implication for this consequential and highly expensive field of public policy is that those legislators, administrators, bankers, media representatives, and members of the public who value honest numbers should not trust the cost estimates presented by infrastructure promoters and forecasters. Another important implication is that institutional checks and balances--including financial, professional, or even criminal penalties for consistent or foreseeable estimation errors--should be developed to ensure the production of less deceptive cost estimates. The work of designing such checks and balances has been begun elsewhere, with a focus on four basic instruments of accountability: (1) increased transparency, (2) the use of performance specifications, (3) explicit formulation of the



regulatory regimes that apply to project development and implementation, and (4) the

involvement of private risk capital, even in public projects (Bruzelius et al., 1998; Flyvbjerg

et al., in press).

## Acknowledgments


The authors wish to thank Martin Wachs, Don Pickrell, and three anonymous JAPA referees

for valuable comments on an earlier draft of the paper. Research for the paper was supported

by the Danish Transport Council and Aalborg University, Denmark.


## Notes

1. Merewitz's (1973a, 1973b) study compared cost overrun in urban rapid transit projects, especially the San

Francisco Bay Area Rapid Transit (BART) system, with overrun in other types of public works projects.

Merewitz's aims were thus different from ours, and his sample of transportation projects was substantially

smaller: 17 rapid transit projects and 49 highway projects, compared with our 58 rail projects, 167 highway

projects, and 33 bridge and tunnel projects. In addition to issues of a small sample, in our attempt to replicate

Merewitz's analysis we found that his handling of data raises a number of other issues. First, Merewitz did not

correct his cost data for inflation, i.e., current prices were used instead of fixed ones. This is known to be a major

source of error due to varying inflation rates between projects and varying duration of construction periods.

Second, in statistical tests, Merewitz compared the mean cost overrun of subgroups of projects (e.g., rapid

transit) with the grand mean of overrun for all projects, thus making the error of comparing projects with

themselves. Subgroups should be tested directly against other subgroups in deciding whether they differ at all

and, if so, which ones differ. Third, Merewitz's two reports (1973a, 1973b) are inconsistent. One (Merewitz,

1973a) calculates the grand mean of cost overrun as the average of means for subgroups; that is, the grand mean

is unweighted, where common practice is to use the weighted mean, as appears to be the approach taken in the

other (Merewitz, 1973b). Fourth, due to insufficient information, the p-values calculated by Merewitz are

difficult to verify; most likely they are flawed, however, and Merewitz's one-sided p-values are misleading.

Finally, Merewitz used a debatable assumption about symmetry, which has more impact for the nonparametric

test used than nonnormality has for parametric methods. Despite these shortcomings, the approach taken in

Merewitz's study was innovative for its time and in principle pointed in the right direction regarding how to



analyze cost escalation in public works projects. The study cannot be said to be a true large-sample study for transportation infrastructure, however, and its statistical significance is unclear.

2. The Mâitre d'Oeuvre was an organization established to monitor project planning and implementation for the Channel tunnel. It was established in 1985, and until 1988 it represented the owners. In 1988 it was reverted to an impartial position (Major Projects Association, 1994, pp. 151-153).

**Appendix**

The first task of the research reported in this paper was to establish a sample of
infrastructure projects substantially larger than what is common in this area of research, a
sample large enough to allow statistical analyses of costs. Here a first problem was that data
on actual costs in transportation infrastructure projects are relatively difficult to come by. One
reason is that it is quite time consuming to produce such data. For public sector projects,
funding and accounting procedures are typically unfit for keeping track of the multiple and
complex changes that occur in total project costs over time. For large projects, the relevant
time frame may cover 5, 10, or more fiscal years from decision to build, until construction
starts, until the project is completed and operations begin. Reconstructing the actual total



costs of a public project, therefore, typically entails long and difficult archival work and complex accounting. For private projects, even if funding and accounting practices may be more conducive to producing data on actual total costs, such data are often classified to keep them from the hands of competitors. Unfortunately, this also tends to keep data from the hands of scholars. And for both public and private projects, data on actual costs may be held back by project owners because more often than not, actual costs reveal substantial cost escalation, and cost escalation is normally considered somewhat of an embarrassment to promoters and owners. In sum, establishing reliable data on actual costs for even a single transportation infrastructure project is often highly timeconsuming or simply impossible.

This state of affairs explains why small-sample studies dominate scholarship in this field of research. But despite the problems mentioned, after 4 years of data collection and refinement, we were able to establish a sample of 258 transportation infrastructure projects with data on both actual construction costs and estimated costs at the time of decision to build. The project portfolio is worth approximately $90 billion (U.S.; 1995 prices). The project types are bridges, tunnels, highways, freeways, high-speed rail, urban rail, and conventional (interurban) rail. The projects are located in 20 countries on 5 continents, including both developed and developing nations. The projects were completed between 1927 and 1998. Older projects were included in the sample in order to test whether the accuracy of estimated costs improve over time. The construction costs of projects range from $1.5 million (U.S.) to $8.5 billion (U.S.; 1995 prices), with the smallest projects typically being stretches of roads in larger road schemes, and the largest projects being rail links, tunnels, and bridges. As far as we know, this is the largest sample of projects with data on cost development that has been established in this field of research.

In statistical analysis, data should be a sample from a larger population, and the sample should represent the population properly. These requirements are ideally satisfied by drawing the sample by randomized lot. Randomization ensures with high probability that



non-controllable factors are equalized. A sample should also be designed such that the representation of subgroups corresponds to their occurrence and importance in the population. In studies of human affairs, however, where controlled laboratory experiments often cannot be conducted, it is frequently impossible to meet these ideal conditions. This is also the case for the current study, and we therefore had to take a different approach to sampling and statistical analysis.

We selected the projects for the sample on the basis of data availability. All projects that we knew of for which data on construction cost development were obtainable were considered for inclusion in the sample. Cost development is defined as the difference between actual and estimated costs in percentage of estimated costs, with all costs measured in fixed prices. Actual costs are defined as real, accounted costs determined at the time of completing a project. Estimated costs are defined as budgeted, or forecast, costs at the time of decision to build. Even if the project planning process varies with project type, country, and time, it is typically possible to locate for a given project a specific point in the process that can be identified as the time of decision to build the project. Usually a cost estimate was available for this point in time. If not, the closest available estimate was used, typically a later estimate resulting in a conservative bias in our measurement of cost development. Cost data were collected from a variety of sources, including annual project accounts, questionnaires, interviews, and other studies.

Data on cost development were available for 343 projects. We then rejected 85 projects because of insufficient data quality. For instance, for some projects we could not obtain a clear answer regarding what was included in costs, or whether cost data were given in current or fixed prices, or which price level (year) had been used in estimating and discounting costs. More specifically, of those 85 projects, we rejected 27 because we could not establish whether or not cost data were valid and reliable. We rejected 12 projects because they had been completed before 1915 and no reliable indices were available for discounting



costs to the present. Finally, we excluded 46 projects because cost development for them turned out to have been calculated before construction was completed and operations begun; therefore, the actual final costs for these projects may be different from the cost estimates used to calculate cost development, and no information was available on actual final costs. In addition to the 85 rejected projects mentioned here, we also rejected a number of projects to avoid double counting of projects. This typically involved projects from other studies that appeared in more than one study or where we had a strong suspicion that this might be the case. In sum, all projects for which data was considered valid and reliable were included in the sample. This covers both projects for which we ourselves collected the data and projects for which other researchers in other studies did the data collection (Fouracre et al., 1990; Hall, 1980; Leavitt et al., 1993; Lewis, 1986; Merewitz, 1973a; National Audit Office & Department of Transport, 1985, 1992; National Audit Office, Department of Transport, Scottish Development Department, & Welsh Office, 1988; Pickrell, 1990; Riksrevisionsverket, 1994; Vejdirektoratet, 1995; Walmsley & Pickett, 1992). Cost data were made comparable across projects by discounting prices to the 1995 level and calculating them in Euros, using the appropriate geographical, sectoral, and historical indices for discounting and the appropriate exchange rates for conversion between currencies.

Our own data collection concentrated on large European projects because too few data existed for this type of project to allow comparative studies. For instance, for projects with actual construction costs larger than 500 million Euros (1995 prices; EUR1=U.S.$1.29 in 1995), we were initially able to identify from other studies only two European projects for which data were available on both actual and estimated costs. If we lowered the project size and looked at projects larger than 100 million Euros, we were able to identify such data for eight European projects. We saw the lack of reliable cost data for European projects as particularly problematic since the Commission of the European Union had just launched its policy for establishing the so-called trans-European transport networks (TTEN), which would



involve the construction of a large number of major transportation infrastructure projects across Europe at an initial cost of 220 billion Euros (Commission of the European Union, 1993, p. 75). As regards costs, we concluded that the knowledge base for the Commission's policy was less than well developed and we hoped to help remedy this situation through our data collection. Our efforts on this point proved successful. We collected primary data on cost for 37 projects in Denmark, France, Germany, Sweden, and the U.K. and were thus able to increase many times the number of large European projects with reliable data for both actual and estimated costs, allowing for the first time a comparative study for this type of project in which statistical methods can be applied.

As for any sample, a key question is whether the sample is representative of the population. Here the question is whether the projects included in the sample are representative of the population of transportation infrastructure projects. Since the criterion for sampling was data availability, this question translates into one of whether projects with available data are representative. There are four reasons why this is probably not the case. First, it may be speculated that projects that are managed well with respect to data availability may also be managed well in other respects, resulting in better than average (i.e., nonrepresentative) performance for such projects. Second, it has been argued that the very existence of data that make the evaluation of performance possible may contribute to improved performance when such data are used by project management to monitor projects (World Bank, 1994, p. 17). Again, such projects would not be representative of the project population. Third, we might speculate that managers of projects with a particularly bad track record regarding cost escalation have an interest in not making cost data available, which would then result in underrepresentation of such projects in the sample. Conversely, managers of projects with a good track record for costs might be interested in making this public, resulting in overrepresentation of these projects. Fourth, and finally, even where managers have made cost data available, they may have chosen to give out data that present



their projects in as favorable a light as possible. Often there are several estimates of costs to choose from and several calculations of actual costs for a given project at a given time. If researchers collect data by means of survey questionnaires, as is often the case, there might be a temptation for managers to choose the combination of actual and estimated costs that suits them best, possibly a combination that makes their projects look good.

The available data do not allow an exact, empirical assessment of the magnitude of the problem of misrepresentation. But the few data that exist that shed light on this problem support the thesis that data are biased. When we compared data from the Swedish Auditor General for a subsample of road projects, for which the problems of misrepresentation did not seem to be an issue, with data for all road projects in our sample, we found that cost escalation in the Swedish subsample is significantly higher than for all projects (Holm, 1999, pp.11-15). We conclude, for the reasons given above, that most likely the sample is biased and the bias is conservative. In other words, the difference between actual and estimated costs estimated from the sample is likely to be lower than the difference in the project population. This should be kept in mind when interpreting the results from statistical analyses of the sample. The sample is not perfect by any means. Still it is the best obtainable sample given the current state of the art in this field of research.

In the statistical analyses, percentage cost development in the sample is considered normally distributed unless otherwise stated. Residual plots, not shown here, indicate that normal distribution might not be completely satisfied, the distributions being somewhat skewed with larger upper tails. However, transformations (e.g., the logarithmic one) do not improve this significantly. For simplicity, therefore, no transformation has been made, unless otherwise stated.

The subdivisions of the sample implemented as part of analyses entail methodological problems of their own. Thus the representation of observations in different combinations of subgroups is quite skewed for the data considered. The analysis would be



improved considerably if the representation were more even. Partial and complete confounding occur; that is if a combination of two or more effects is significant, it is sometimes difficult to decide whether one or the other or both, cause the difference. For interactions, often not all the combinations are represented or the representations can be quite scarce. We have adapted our interpretations of the data to these limitations, needless to say. If better data could be gathered, sharper conclusions could be made.

The statistical models used are linear normal models (i.e., analysis of variance and regression analysis with the appropriate F-tests and t-tests). The tests of hypotheses concerning mean values are known to be robust to deviations from normality. Also, chi-square tests for independence have been used for count data. For each test, the p-value has been reported. This value is a measure for rareness if identity of groups is assumed. Traditionally, a p-value less than 0.01 is considered highly significant and less than 0.05 significant, whereas a larger p-value means that the deviation could be due do chance.



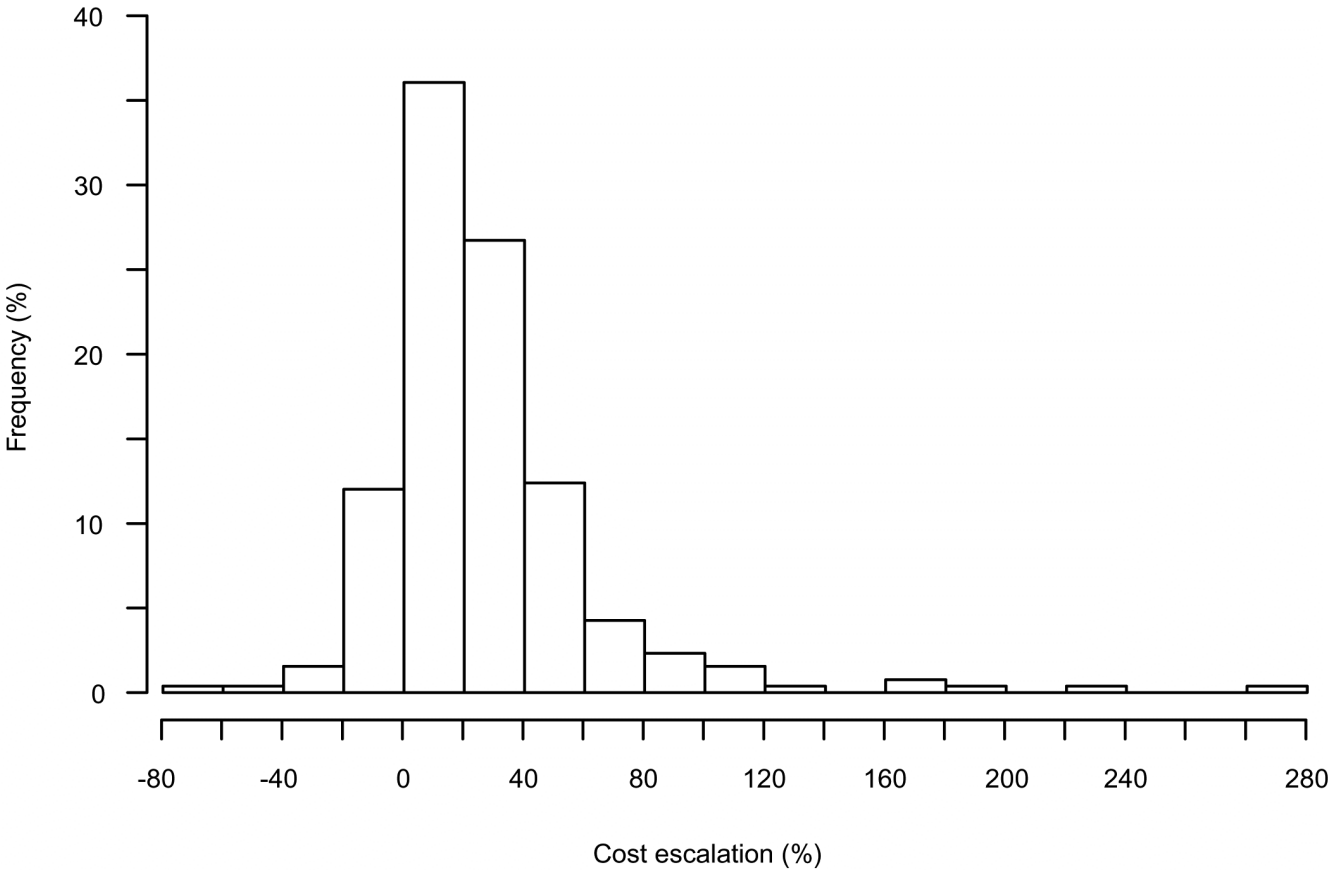

**Figure** 1. Inaccuracy of cost estimates in 258 transportation infrastructure projects (fixed prices).



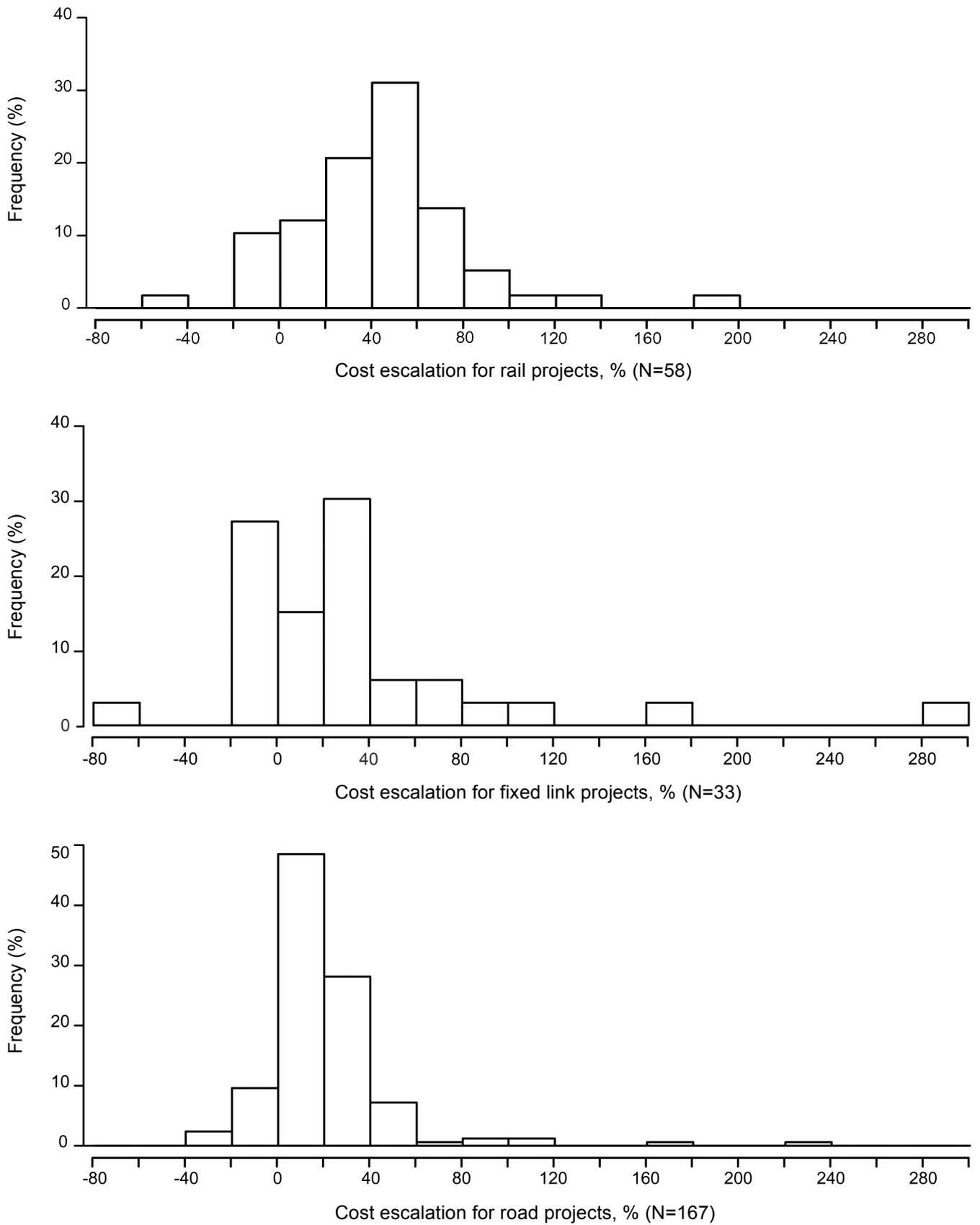

**Figure** 2. Inaccuracy of cost estimates in rail, fixed link, and road projects (fixed prices).



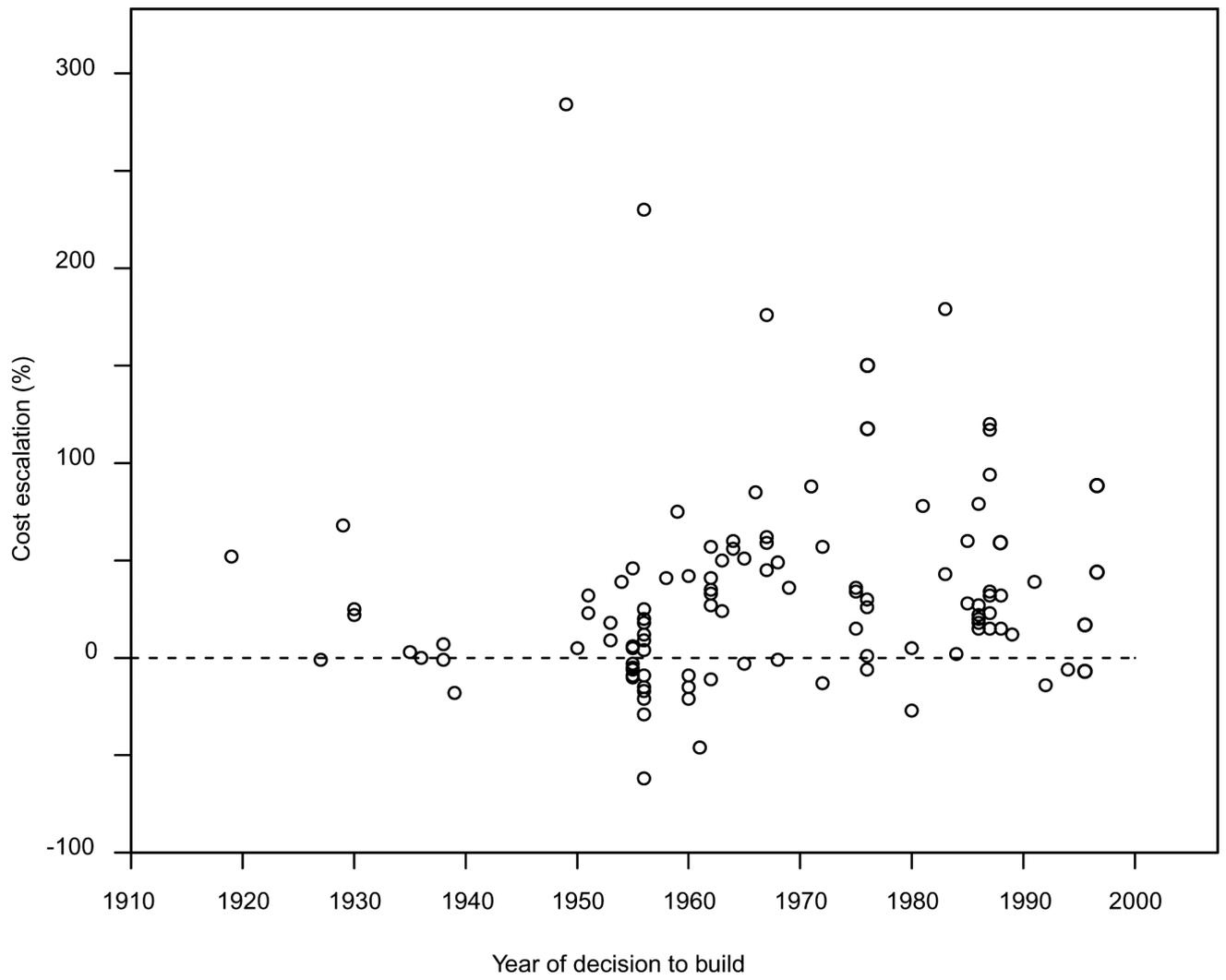

**Figure** 3. Inaccuracy of cost estimates in transportation projects over time, 1910-1998 (fixed

prices, 111 projects)



**Table** 1. Inaccuracy of transportation project cost estimates by type of project (fixed prices)

| Type of project | Number of cases (N) | Average cost escalation (%) | Standard deviation | Level of significance, (p) |
|---|---|---|---|---|
| Rail | 58 | 44.7 | 38.4 | <0.001 |
| Fixed links | 33 | 33.8 | 62.4 | 0.004 |
| Roads | 167 | 20.4 | 29.9 | <0.001 |
| All projects | 258 | 27.6 | 38.7 | <0.001 |



**Table** 2. Inaccuracy of transportation project cost estimates by geographic location (fixed prices)

| Type of project | Europe | | | North America | | | Other geographical areas | | |
|---|---|---|---|---|---|---|---|---|---|
| | Number of projects (N) | Average cost escalation (%) | Standard deviation | Number of projects (N) | Average cost escalation (%) | Standard deviation | Number of projects (N) | Average cost escalation (%) | Standard deviation |
| Rail | 23 | 34.2 | 25.1 | 19 | 40.8 | 36.8 | 16 | 64.6 | 49.5 |
| Fixed links | 15 | 43.4 | 52.0 | 18 | 25.7 | 70.5 | 0 | - | - |
| Roads | 143 | 22.4 | 24.9 | 24 | 8.4 | 49.4 | 0 | - | - |
| Total | 181 | 25.7 | 28.7 | 61 | 23.6 | 54.2 | 16 | 64.6 | 49.5 |